%% file: main.tex
\documentclass[aps,prd,showpacs,floatfix,twocolumn,superscriptaddress,nofootinbib,preprintnumbers,
10pt]{revtex4-2}
\input{aux_files/mycommands}

\begin{document}

\preprint{MIT-CTP/5956}

\title{Bridging Simulations and EFT: A Hybrid Model of the
Lyman-$\alpha$ Forest Field}

\author{Roger de Belsunce}
\email{belsunce@mit.edu}
\affiliation{\MIT}\affiliation{\IAIFI}
\author{Boryana Hadzhiyska}
\email{bth26@cam.ac.uk}
\affiliation{\Kavli}
\author{Mikhail M. Ivanov}
\email{ivanov99@mit.edu}
\affiliation{\MIT}\affiliation{\IAIFI}

\begin{abstract}
The Lyman-$\alpha$ (\Lya) forest is a unique probe of cosmology and
the intergalactic medium at high redshift and small scales. The
statistical power of the ongoing Dark Energy Spectroscopic
Instrument (DESI) demands precise theoretical tools to model the
\Lya forest. We present a hybrid effective field theory (HEFT)
forward model in redshift space that leverages the accuracy of
non-linear 
particle displacements computed using the $N$-body simulation suite
\textsc{AbacusSummit} with the predictive power of an analytical,
perturbative bias forward model in the framework of the effective
field theory (EFT). 
The residual noise
between the model 
and the simulated 
\Lya field
has a nearly white (scale-
and orientation-independent) 
power spectrum on quasi-linear scales,  
substantially simplifying
its modeling compared 
to a purely 
perturbative description.
As a consequence
of the improved control over the 3D \Lya forest stochasticity, we find
agreement between the modeled and the  true power spectra at the
5\% level down to scales of $\kmax \simlt 1 \hMpcinv$. This procedure offers a promising path toward constructing efficient and accurate emulators to predict
large-scale clustering summary statistics for full-shape cosmological analyses of \Lya forest
data from both DESI and its successor, DESI-II.
\end{abstract}

\maketitle

\section{Introduction}
Light emitted by high-redshift quasars ($2\simlt z \simlt 5$) is
absorbed in the low-density, highly ionized intergalactic medium
(IGM). This creates a series of absorption features in their observed
spectrum -- the Lyman-$\alpha$ (\Lya) forest
\cite{2016ARA&A..54..313M}. Correlations along the line-of-sight have
provided strong constraints on small-scale physics using  the
one-dimensional power spectrum  \citep[see, e.g.,][]{
  Viel:2005,McDonald06, Chabanier:2019, Pedersen:2020,
2023MNRAS.526.5118R, 2024MNRAS.tmp..176K} through its sensitivity to
the neutrino masses \citep{Seljak:2005, Viel:2010, PYB13,
Palanque2020}, primordial black holes \citep{Afshordi:2003,
Murgia:2019, Ivanov:2025pbu}, and -- at even smaller scales -- dark
matter models \citep{Viel:2013, Baur:2016, Irsic17, Kobayashi:2017,
  Armengaud:2017, Murgia:2018,Garzilli:2019, Irsic:2020, Rogers:2022,
Villasenor:2023, Irsic:2023}, early dark energy models
\citep{2023PhRvL.131t1001G}, and thermal properties of the ionized
(cold) IGM \citep{Zaldarriaga:2002, Meiksin:2009,McQuinn:2016,
  Viel:2006, Walther:2019, Bolton:2008, Garzilli:2012, Gaikwad:2019,
Boera:2019, Gaikwad:2021, Wilson:2022, Villasenor:2022}.
Correlating different lines of sight serves as a large-scale
structure tracer, capable of constraining the expansion history of
the Universe through measurements of the baryonic acoustic oscillations
\citep[BAO;][]{McDonald:2007,Slosar2013, Busca:2013, dMdB:2020,
DESI_lya_2024, DESI_BAO_2024}, and the broadband shape of the
(large-scale) 3D \Lya correlation function \citep{Slosar2013,
Cuceu:2021, Cuceu:2023, Gordon:2023, Cuceu:2025nvl}.

A central assumption underlying the use of the \Lya forest as a small-scale tracer is that hydrodynamic simulations can accurately capture the relevant small-scale physics (see, e.g.,~\cite{Bolton:2016bfs,Lukic:2015,
Chabanier:2024knr}).
Whilst heuristic fitting
functions calibrated on hydrodynamic simulations \cite{McDonald:2001,
Arinyo-i-Prats:2015, DESI_lya_2024} have resulted in robust
cosmological constraints of the currently observing Dark Energy
Spectroscopic Instrument (DESI; \cite{DESI:2016, DESI_BAO_2024})
year-1 data \cite{DESI_lya_2024}, DESI is forecasted to reach a
cumulative precision below 0.2\% over its
five-year survey combining all tracers and redshifts
\cite{DESI:2016}. This requires exquisite knowledge over theoretical
systematics and precise simulations. The key quantity of interest for
DESI is to measure the expansion history of our Universe through the
BAO feature. Whilst this oscillatory feature in the power spectrum
(or peak in the correlation function) is remarkably robust,
non-linearities in clustering modify it in two significant ways: the
amplitude of the oscillatory linear signal is damped towards small
scales by incoherent bulk motions on the BAO scale and out-of-phase
contributions from clustering induce a shift in the observed BAO
position (see, e.g.,~\cite{deBelsunce:2024rvv} in the context of the
\Lya forest). 
In particular, the
phenomenological fitting
model employed for current cosmological \LyaF analyses has been shown to bias
the 
inference of the BAO scaling parameters \cite{deBelsunce:2024rvv, Abacus_BAO_Lya:2025} 
at the $\sim 0.3\%$
level, which matches the 
expected statistical
errors of DESI year-5
data~\cite{Aghamousa:2016zmz}. 
Accounting for this systematic error
will substantially inflate 
the error budget.

The bias in the BAO measurements can be mitigated within the
framework of the effective field theory (EFT) of large-scale
structure \cite{Ivanov:2023yla, Abacus_BAO_Lya:2025}. The EFT program
relies strictly on the symmetries pertaining to the modeled tracer
(here; the \Lya forest with translational invariance, $SO(2)$
  rotation invariance
  around the line-of-sight, sightline reflection symmetry, and the
equivalence principle) as well as dimensional analysis
\cite{McDonald:2009dh, Baumann:2010tm, Carrasco:2012cv,Ivanov:2023yla,
deBelsunce:2025edy}. Yet, the theory modeling at small scales is affected by non-Poisson stochasticity  \cite{deBelsunce:2025bqc} (the analog of the one-halo term in the context of galaxy clustering), whose 
spectrum exhibits
a noticeable scale 
and direction 
dependence 
for wavenumbers
greater than $0.8~h\text{Mpc}^{-1}$.
Ultimately, this stochastic 
noise sets the limit
of the applicability
of EFT~\cite{Baldauf:2015tla,Baldauf:2015zga,Schmittfull:2018yuk,Schmittfull:2020trd}.

In the present work, we increase the  reach of EFT by addressing this small-scale modeling challenge of the \LyaF. Therefore, we leverage the
accuracy of $N$-body simulations
for computing particle displacements \cite{2020MNRAS.492.5754M} and combine it
with the perturbative forward model of the \Lya forest at the
field-level presented in Ref.~\cite{deBelsunce:2025bqc}. 
To validate our hybrid model, we evolve a set of cosmological initial conditions and fit a bias expansion at the field level. The resulting fields are then compared to the input simulations at the field level and using the power spectrum. This approach benefits from cosmic variance cancellation when choosing the same set of initial conditions for the forward model and the simulations. Additionally, it enables generating simulations at cosmological volumes whilst being calibrated on input simulations that, e.g.,~resolve features  affecting the BAO feature at a fraction of the computational cost. Further, we outline a path to emulate summary statistics replacing current fitting functions \citep{DESI_lya_2024}
 used for full-shape
cosmological analyses of \Lya forest data from DESI. This will enable analyses of the
broadband shape of the large-scale 3D correlation function
\cite{Cuceu:2025nvl} (or its Fourier counterpart, the 3D power
  spectrum \cite{Font-Ribera:2018, deBelsunce:2024knf,
Horowitz:2024nny}) and even compressed 3D bispectrum
statistics \cite{deBelsunce:2025edy}. With a number of near-future
surveys  that will capture spectra in both high and medium resolution
bands, such as the WEAVE-QSO survey \citep{2016sf2a.conf..259P}, the
Prime Focus Spectrograph \citep[PFS;][]{2022PFSGE} and 4MOST
\citep{2019Msngr.175....3D}, this toolkit is timely for the
cosmological analysis of the \Lya forest.

This paper is organized as follows: In
Sec.~\ref{sec:LyaHEFT} we briefly summarize the hybrid effective
field theory model and the bias expansion of the \Lya forest in Lagrangian perturbation theory. We
present the employed \abacus simulations of the \Lya forest in
Sec.~\ref{sec:simulations}. In Sec.~\ref{sec:res} we present our
results and compare the \Lya forest fields from \abacus to the ones obtained from our forward model by evolving the same set of initial conditions and fitting them to the input simulation. We present our
conclusions in Sec.~\ref{sec:conclusion} and discuss extensions to construct an emulator for full-shape cosmological 
analyses of the \Lya forest including cross-correlation with 
biased tracers such as high-redshift galaxies or quasars. 

\section{Hybrid EFT Forward Model} \label{sec:LyaHEFT}
The hybrid effective field theory (HEFT; \cite{2020MNRAS.492.5754M})
combines a perturbative bias expansion using Lagrangian
perturbation theory (LPT) with $N$-body techniques to solve the full
non-linear dynamics to compute particle displacements. In the
following, we will briefly summarize HEFT and refer the reader to
Refs.~\cite{2002PhR...367....1B,2018PhR...733....1D}
for a review of perturbation theory with particular emphasis to the
Lagrangian framework \cite{2014JCAP...05..022P, 2015JCAP...09..014V,
Chen:2020fxs, Chen:2020zjt}. LPT is described by infinitesimal fluid elements
with initial (Lagrangian) positions $\bm{q}$. Now, the dynamics are
described by the displacement $\bm{\Psi}(\bm{q},\eta)$ which are, in
turn, generated by the gravitational potential. The corresponding
Eulerian (comoving) positions $\bm{x}$ of the fluid element at some
conformal time $\eta$ are given by $\bm{x}(\bm{q},\eta) = \bm{q} +
\bm{\Psi}(\bm{q},\eta)$ \cite{2015JCAP...09..014V,2014JCAP...05..022P}.
HEFT  has been shown to perform very well
in both simulations and observations
\citep{2020MNRAS.492.5754M,2021MNRAS.505.1422K,2021JCAP...09..020H,2023MNRAS.524.2407Z,2023MNRAS.520.3725P}.

\subsection{HEFT Operators}\label{sec:HEFT_operators}
We write the dependence of
the \Lya forest field along its trajectory as an expansion
to second order in the initial conditions
\cite{McDonald:2009dh,Baumann:2010tm, Carrasco:2012cv,
Ivanov:2023yla,Sullivan_Chen_LPNG_HEFT, deBelsunce:2025edy}:
\begin{align}
  \label{eq:bias_expansion}
  F(\bm{q}) &= \beta_{cb} 1(\bm{q}) + \beta_1 \delta_L(\bm{q}) + \beta_2
  \big(\delta_L^2(\bm{q})-\langle\delta_L^2\rangle\big) \\ \nonumber
  &\quad + \beta_s \big(s_L^2(\bm{q})-\langle s_L^2\rangle \big) + \beta_\nabla
  \nabla^2 \delta_L(\bm{q}) \\ \nonumber 
  &\quad +
  \beta_\eta \big(\eta(\bm{q})-\langle \eta\rangle \big)(\bm{q}) + \beta_{\delta\eta} (\delta \eta)(\bm{q}) \nonumber \\  
  &\quad + \beta_{\eta^2} (\eta^2(\bm{q})  - \langle \eta^2
  \rangle)(\bm{q}) + \beta_{\td^3}
  \td^3(\bm{q}) \nonumber \\ 
  &\quad+ \beta_{KK_\parallel} \big(KK_\parallel(\bm{q})-\langle KK_\parallel\rangle \big)(\bm{q}) \nonumber \,,
\end{align}
where $\beta_{cb}$, $\beta_1$, $\beta_2$, $\beta_s$, $\beta_\nabla$, $\beta_\eta$, $\beta_{\delta\eta}$,
$\beta_{\eta^2}$,
$\beta_{\td^3}$, and $\beta_{KK_\parallel}$ are free bias transfer functions
of Fourier wavenumber $k$ and angle to the line-of-sight, $\mu =
k_\parallel / k$, $\langle \dots \rangle$ denotes the ensemble average, and $\td_L=D(z)\td_{L}^{(z=0)}$ is the linear overdensity which is evolved by the linear growth factor $D$. Due to
the degeneracy between the operators $KK_\parallel$ and
$\Pi_\parallel^{[2]}$ 
in the transfer function approach
\cite{deBelsunce:2025bqc}, we
remove the latter and remap the redshift space distortion parameter
to be $\mathcal{O}_\eta(\vk,\mu) \equiv \delta_m(\vk) - 3/7
\ \mu^2 s^2(\vk)$, following Ref.~\cite{deBelsunce:2025bqc}. We
also add a transfer function for the non-linear redshift space distortions $\beta_{cb}$ and use the renormalized cubic operator
$\delta_1^3\rightarrow\delta_1^3-3\sigma^2\delta_1$ where $\sigma^2$
is the mean-subtracted mean of the squared field $\delta_1$. The tidal field
is given by $s_L^2 = s_{ij} s^{ij}$ with the traceless tidal tensor $s_{ij}\equiv ( \partial_i
\partial_j/\partial^2 - \delta_{ij}/3 )\ \delta_L$, and
$\nabla^2\delta_L$ is  the lowest-order non-local term. In contrast to
galaxies, the \LyaF has additional operators acting along the
line-of-sight due to a relaxed set of symmetries. We use the same \Lya-specific
bias operators as Ref.~\cite{deBelsunce:2025bqc}:
\begin{equation}
  \eta = \hat z^i \hat z^j \partial_i \partial_j/\partial^2 \delta_L
  = \left(\partial_z \partial_z/\partial^2 \right)\delta_L\,,
\end{equation}
with
\begin{equation}
  KK_\parallel = \hat z^i \hat z^l s_{ij} s^{j}_{l} = s_{zi} s^{iz}\,,
\end{equation}
where we assume the line-of-sight to be in the $\hat z$ direction. Each particle in the $N$-body simulation is associated with the weight given by Eq.~\eqref{eq:bias_expansion}.

The functional can then be advected to  redshift-space (Eulerian)
position $\bm{x}$ \cite{2008PhRvD..78h3519M}
\begin{equation}
  1 + \delta_{\rm F}^{(s)}(\bm{x})
  = \int d^3\bm{q}\,F(\bm{q})\,
  \delta^D(\bm{x}-\bm{q}-\bm{\Psi^{(s)}}(\bm{q}))\,.
  \label{eqn:deltag}
\end{equation}
which is done numerically by using the final
positions of the dark matter particles in the simulation. The Kronecker delta and Lagrangian displacement vector are denoted by $\td^{\rm D}$  and  $\bm{\Psi}$, respectively. In summary, the key
difference to EFT is that in the present work we use the cubic operator $\td^3$ and the
non-linear displacements of the particles obtained from the $N$-body
simulations and in EFT we use Zel'dovich displacements, $\delta_Z$.


\subsection{Redshift-Space Advection}


The Lyman-$\alpha$ forest is inherently a line-of-sight observable, and thus any forward-modeling framework must accurately incorporate redshift-space effects. A key advantage of the HEFT approach is that redshift-space distortions can be implemented simply by replacing the real-space displacement field, $\boldsymbol{\Psi}(\bm{q})$, with its redshift-space counterpart, $\boldsymbol{\Psi}^{(s)}(\bm{q})$, when advecting the operators. The latter can be written as the sum of the full nonlinear real-space displacement and a line-of-sight (LOS) velocity-induced component. For the real-space displacement, we follow the standard HEFT procedure:\, for each particle, we compute the nonlinear displacement as the difference between its final Eulerian position at $z=2.5$ and its Lagrangian coordinate inferred from the stored initial conditions, $\bm{q}$.\footnote{Note that because \abacus records the initial positions at reduced precision, this step introduces some stochastic noise into the HEFT operators.}

Once the initial coordinate $\bm{q}$ is known, we assign the values of each HEFT operator $\mathcal{O}(\bm{q})$ by mapping $\bm{q}$ to its nearest-grid-point (NGP) index on the initial grid. These operators are computed using the full $2304^3$ linear density field (matching the resolution of the input initial conditions (ICs):\, the native simulation resolution is $6912^3$, but using it directly would be computationally prohibitive). The NGP assignment step is an additional source of small-scale noise. The advected operators can in principle be reconstructed at any mesh size:\, the optimal choice is to match the IC resolution ($2304^3$), which preserves the smallest relevant scales. However, since our analysis only requires modes up to $k \approx 2\,h\,\mathrm{Mpc}^{-1}$ (appropriate for DESI-like surveys), and for numerical stability, we adopt $1152^3$ as our working resolution. To further suppress noise and, since we are dominated by the NGP step and the limited-precision IC positions, we advect all $6912^3$ simulation particles at $z=2.5$, which reduces the effective operator noise by a factor of a few on large scales (bringing it close to the noise floor of EFT) and yields noticeably improved performance relative to EFT on small scales.

We now comment on how to construct the redshift-space displacement, i.e.~the velocity-generated contribution added on top of the nonlinear real-space displacement. We explored several choices:

{\bf (i) Zel'dovich velocity field} \\
Our fiducial model uses the Zel'dovich prediction for the LOS velocity displacements, $f_1 D_1\,\boldsymbol{\Psi}_1$ which is  added  to the full non-linear real-space displacements. This prescription performs well and provides stable behavior across all scales of interest.

{\bf (ii) 2LPT velocity field} \\
We also implemented the full second order LPT (2LPT) velocity displacements, $f_1 D_1\,\boldsymbol{\Psi}_1 + f_2 D_2\,\boldsymbol{\Psi}_2$, computed using a dedicated pipeline with $3/2$-zero-padding and meshes of size  $2304^3$ to preserve the original IC resolution without introducing additional smoothing. 
We find no measurable improvement using 2LPT in lieu of the Zel’dovich-only model. There are two reasons for this: First, LPT lacks the stochastic component of the velocity field, which dominates the large-scale noise budget, so deterministic corrections beyond first order cannot resolve the missing-velocity problem. Second, any potential improvement from 2LPT is absorbed by the set of operators and transfer functions already included in the model.

{\bf (iii) Raw simulation velocities} \\
We attempted to use the particle velocities directly, converting them to displacements via $v/(aH)$. Adding this to the nonlinear real-space displacement yields the exact non-linear redshift-space displacement. However, this choice performed poorly on small scales as the raw velocities introduce a significant amount of small-scale noise and multi-streaming structure that is not compatible with the smooth HEFT operator fields. Additionally, we attempted to smooth the velocity field constructed from the raw particle velocities, suppress the small scale noisy modes with a smooth tanh($\cdot$) taper, varying the effective maximum mode, $k_{\rm max}$, and sample it at the location of the particles before performing the advection in the line-of-sight direction for the HEFT operators. Although we tested several values for $k_{\rm max} = 0.5, \ 1, \ 2, \ 4 \ h/{\rm Mpc}$, this approach yielded worse performance on our two most important metrics: cross-correlation coefficient, $r(k, \ \mu)$, and error power spectrum, $P_{\rm err}(k, \ \mu)$. 

{\bf (iv) Aspirational approaches} \\
A more accurate estimate of the velocity displacement could in principle be obtained by finite-differencing nonlinear displacements from snapshots at two nearby redshifts. However, \abacus does not provide snapshots sufficiently close to $z=2.5$ (the nearest are at $z=2$ and $z=3$), and the numerical derivative would likely introduce significant noise. 

We proceed with the Zel'dovich  advected displacements as our
baseline modeling
choice for the redshift space distortions.

\subsection{Transfer Function Fitting Procedure}

We fit for the bias transfer functions by  minimizing the
difference between the forward model, introduced in Sec.~\ref{sec:HEFT_operators} and the observed transmitted flux
fraction obtained from the \abacus simulation: $ \td_F = F/\overline{F}(z)-1$ where  $\overline{F}(z)$ is the mean value of
transmission at the redshift of the simulation ($z=2.5$). This
is similar to the procedure adopted in \citep{Schmittfull:2018yuk,
2022MNRAS.514.2198K, Baradaran:2024jlh}, which we follow closely. In
brief, the stochastic residual field $\epsilon(\vk)$, after
removing deterministic contributions, \textit{i.e.}~the cubic bias
expansion, is given by
\begin{equation}
  \epsilon(\vk) = \delta_{F}(\vk) -
  \delta_m(\vk) - \sum_i \beta_i \mathcal{O}_i(\vk)\,,
\end{equation}
for Fourier wavevector $\vk$ and its power spectrum,
referred to as
the error power spectrum is given by
\be
P_{\rm err} (k,\mu) \equiv \langle |\delta^{\mathrm{truth}}_F(\k) -
\delta^{\mathrm{model}}_F(\k)|^2 \rangle\,,
\ee
and reflects the agreement at the level of the phases and will be used as
a key metric to evaluate the performance of our forward model. The
transmitted flux fraction obtained from the simulations is denoted as ``truth.''
These expressions are evaluated at the redshift of the simulation (here: $z=2.5$) but we will suppress the explicit time-dependence throughout this work. 


To determine the value of the transfer functions, we minimize
  the low-pass filtered stochastic field up to some maximum wavenumber,
$k_{\rm max}$, with the filtering removing small-scale modes. The resulting loss function is given by (see, e.g.,~\cite{2022MNRAS.514.2198K})
\begin{equation}
  \int_{k_{\rm bin}, \ \mu_{\rm bin}} \frac{d^3 k}{(2 \pi)^3}
  ||\epsilon(\vk)||^2\,,
\end{equation}
where we define $k_{\rm bin}$ and $\mu_{\rm bin}$ as $k_{\rm min} <
|\vk| < k_{\rm max}, \ \mu_{\rm min} < \mu < \mu_{\rm max}$,
respectively. The solution of the above equation is
\begin{equation} \label{eq:Mij}
  \hat{\beta}_i = M_{ij}^{-1} A_j\,.
\end{equation}
where $A_j$ and $M_{ij}$ are defined as\footnote{Note that here we
  solve the inverse problem given in Eq.~\eqref{eq:Mij} and in
  Ref.~\cite{deBelsunce:2025bqc} we use the ``shifted operator basis''
  and apply the Gram-Schmidt orthogonalization procedure -- we verified
that our results are robust to details of the implementation.}
\begin{eqnarray}
  A_j = \left\langle [\mathcal{O}_j(\mathbf{x}) (\delta_{\rm
  g}(\mathbf{x}) - \delta_m(\mathbf{x}))]_{k_{\rm bin}, \ \mu_{\rm
  bin}} \right\rangle \,,
  \\ \nonumber
  = \int_{k_{\rm bin}, \ \mu_{\rm bin}}
  \frac{d^3k}{(2\pi)^3} \mathcal{O}_j(\vk) [\delta_{\rm g} -
  \delta_m]^*(\vk) \,,
\end{eqnarray}
and
\begin{eqnarray}
  M_{ij} = \left\langle [\mathcal{O}_i(\mathbf{x})
  \mathcal{O}_j(\mathbf{x})]_{k_{\rm bin}, \ \mu_{\rm bin}} \right\rangle ,
  \\ \nonumber
  = \int_{k_{\rm bin}, \ \mu_{\rm bin}} \frac{d^3k}{(2\pi)^3}
  \mathcal{O}_i(\vk) \mathcal{O}_j^*(\vk).
\end{eqnarray}
The fitting is performed using $\Delta\mu=0.1$ and $\Delta
k=0.005\hMpcinv$ and we show the final results in three angular bins for
ease of comparison with Ref.~\cite{deBelsunce:2025bqc}. We find that the final performance is largely insensitive to the choice of $\Delta \mu$ and $\Delta k$.

\section{simulations} \label{sec:simulations}
We validate our perturbative forward model using the \abacus
simulations and compute the displacements using the same simulations.
In brief, \abacus is a suite of cosmological $N$-body simulations
developed for the cosmological analysis of the DESI survey  \citep{2021MNRAS.508.4017M}. The simulations were
run with \textsc{Abacus} \citep{Garrison:2021lfa, Garrison:2018juw},
a high-accuracy, high-performance cosmological $N$-body simulation
code, optimized for GPU architectures and for large-volume
simulations, on the Summit supercomputer at the Oak Ridge Leadership
Computing Facility. We use the \texttt{base} resolution boxes of
\textsc{AbacusSummit}, each of which contains 6912$^3$ particles in a
$2\hinvGpc$ box, each with a mass of $M_{\rm part} = 2.1 \times
10^9\msunoh$. While the \textsc{AbacusSummit} suite spans a wide
range of cosmologies, here we focus on a single realization of the
fiducial cosmology (\textit{Planck} 2018: $\Omega_b h^2 = 0.02237$,
  $\Omega_c h^2 = 0.12$, $h = 0.6736$, $10^9 A_s = 2.0830$, $n_s =
0.9649$, $w_0 = -1$, $w_a = 0$), denoted by \texttt{AbacusSummit\_c000\_ph000}.

\subsection{Lyman-$\alpha$ Forest Mocks}
\label{sec:lya_mocks}
We use synthetic \Lya forest data painted on top of the dark matter field of the
\abacus $N$-body simulation suite
\cite{Hadzhiyska:2023, Abacus_BAO_Lya:2025} which have been developed
to support full-shape analyses of the \Lya forest power spectrum for the
DESI analysis. We briefly summarize the mocks in the
following and refer the reader to Refs.~\cite{Hadzhiyska:2023,
Abacus_BAO_Lya:2025} for a fuller presentation.
The mocks are constructed by mapping gas properties onto the dark
matter field of the \abacus $N$-body
simulations through the Fluctuating Gunn--Peterson
Approximation (FGPA; \cite{Croft98}). This prescription is
calibrated using Ly-$\alpha$ skewers extracted from the
\textsc{IllustrisTNG} hydrodynamical simulations \cite{2022ApJ...930..109Q}.
Each mock volume contains $6912^3$ grid cells, corresponding to a
mean interparticle spacing of
$0.29\,h^{-1}\,\mathrm{Mpc}$. This resolution is comparable to the
Jeans scale at the relevant redshifts ($\sim100\,\mathrm{kpc}/h$).
This finite resolution, however, entails that modes on scales smaller than
the resolution scale are not captured. Therefore the density field is
augmented with
additional small-scale fluctuations to restore the missing power,
i.e.~matching the
one-dimensional power spectrum from \textsc{IllustrisTNG}
\cite{2022ApJ...930..109Q}.

The optical depth field is obtained by converting the dark matter
overdensity into a neutral hydrogen density. Two slightly different
methods are implemented. The first follows the standard FGPA
procedure, while the second introduces a refinement that assigns
redshift-space weights to particles when constructing the optical
depth field directly in redshift space. In contrast, the first
approach builds the optical depth from the real-space density field
and applies redshift-space distortions (RSDs) afterwards (see
\cite{Hadzhiyska:2023} for details). The
resulting optical depth fields are then transformed into transmission
flux spectra. In total we obtain four different  \LyaF mocks stemming
from small differences
in implementing the FGPA procedure, and thus, resulting in different bias
parameters, tabulated in Table 4 in Ref.~\cite{Abacus_BAO_Lya:2025}
using linear theory
and fitted using the one-loop power spectrum in EFT in
\citep{Abacus_BAO_Lya:2025} and tabulated in
Tab.~\ref{tab:abacus_models}. For our baseline results we will use
models `one' and `three' which
translate to a scenario with a low and high redshift space distortion bias parameter ($b_{\eta}$) respectively.
Note that model three is closest to observed \Lya data for linear
bias parameters $b_1$ and $b_\eta$.

Although the FGPA-based models described in \cite{Hadzhiyska:2023}
are simplified, they provide an efficient and transparent link
between the dark matter and neutral hydrogen distributions. More
sophisticated alternatives include the \Lya Mass Association Scheme
(LyMAS; \cite{Peirani:2014,Peirani:2022}), the Iteratively Matched
Statistics method (IMS; \cite{Sorini:2016}), Hydro-BAM
\citep{2022ApJ...927..230S}, cosmic-web-dependent FGPA
\citep{2024A&A...682A..21S}, and (promising) deep-learning
reconstruction methods
\cite{Horowitz:2021olb, Harrington:2021srm, Jacobus:2024yev,
Horowitz:2025rke, Hafezianzadeh:2025ifw}. These extend this framework using
higher-resolution hydrodynamic simulations or machine-learning
techniques to reproduce the Ly-$\alpha$ forest probability
distribution and power spectrum with greater fidelity.

\section{Results}
\label{sec:res}

\begin{table}
  \centering
  \begin{tabular}{c|cccccccc}
    \hline
    \hline
    \vspace{-2ex} \\
    M. & $b_1$ & $b_\eta$ & $b_2$ & $b_{\mathcal{G}_2}$ & $b_{\delta
    \eta}$ & $b_{\eta^2}$ & $b_{(KK)_\parallel}$ &
    $b_{\Pi^{[2]}_\parallel}$ \\ [2ex]
    \hline \vspace{-2ex} \\
    I   & -0.149 &  0.142 & -0.123 & -0.085 & -0.109 & -0.377 &
    -0.075 & -0.305 \\[1ex]
    II  & -0.131 &  0.130 & -0.332 & -0.277 & -0.263 & -0.351 &
    \phantom{-}0.484 & -0.294 \\[1ex]
    III & -0.134 &  0.271 & -0.035 & -0.038 & -0.081 & -0.063 &
    -0.016 & -0.243 \\[1ex]
    IV  & -0.130 &  0.304 & -0.040 & -0.032 & -0.039 & -0.017 &
    0.033 & -0.258 \\[1ex]
    \hline
  \end{tabular}
  \caption{\textbf{\abacus Bias Parameters:} Mean best-fit values for the one-loop EFT parameters
    obtained from the \textsc{AbacusSummit} for each of the models
    (M. I--IV) one to four averaging over twelve simulations for two
    line-of-sights. See Ref.~\cite{Abacus_BAO_Lya:2025} for a
    detailed discussion and presentation of the fitting methodology
    and Ref.~\cite{Hadzhiyska:2023} on specifics how \Lya forests are
    planted onto the $N$-body simulation.
    We choose models I and III for this work as the linear bias
    parameters ($b_1$ and $b_\eta$) are closest to measurements from
    hydrodynamic simulations \cite{Chabanier:2024knr} and DESI data
  \cite{DESI_lya_2024} and quote them here for completeness.}
  \label{tab:abacus_models}
\end{table}

\begin{figure*}
\centering
\includegraphics[width=0.49\linewidth]{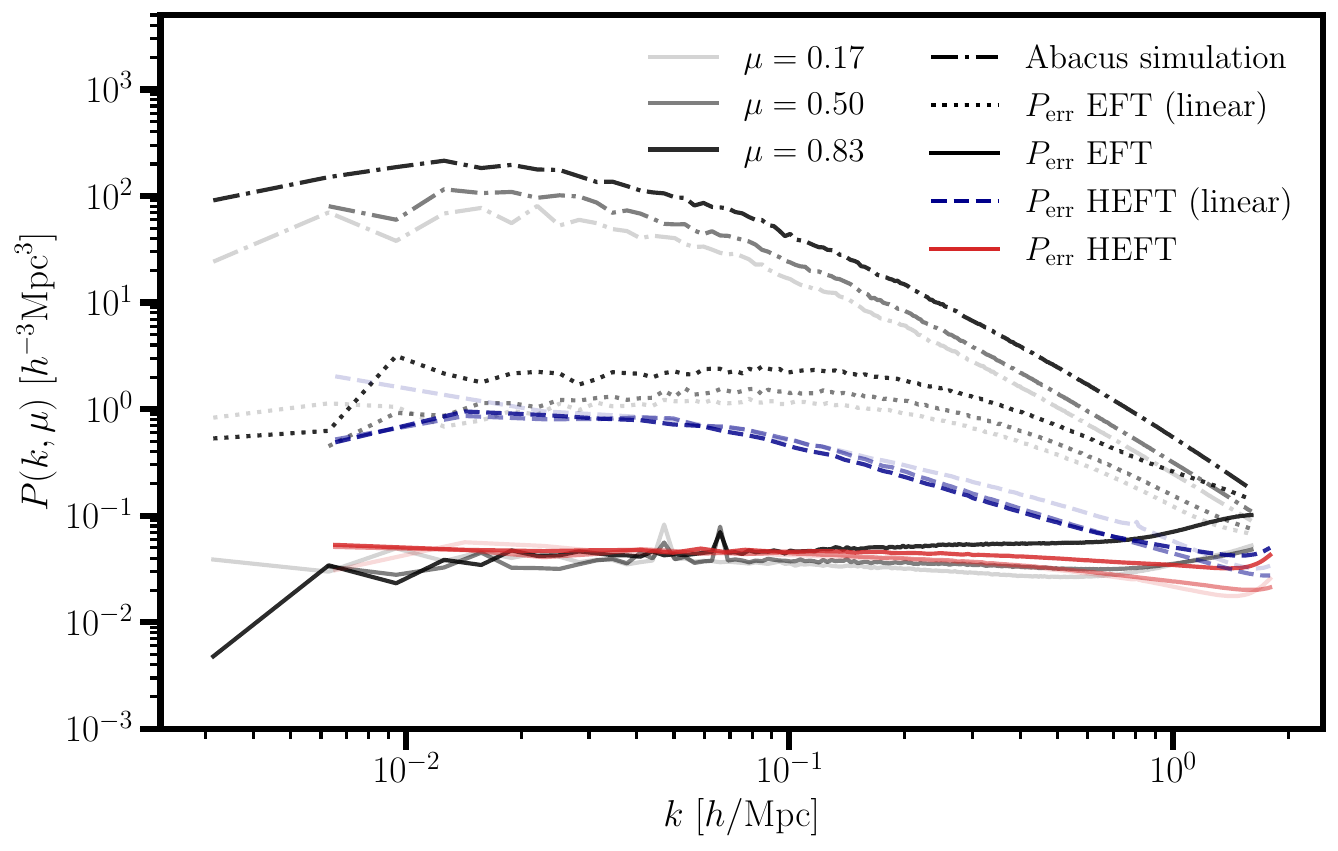}\hfill
\includegraphics[width=0.49\linewidth]{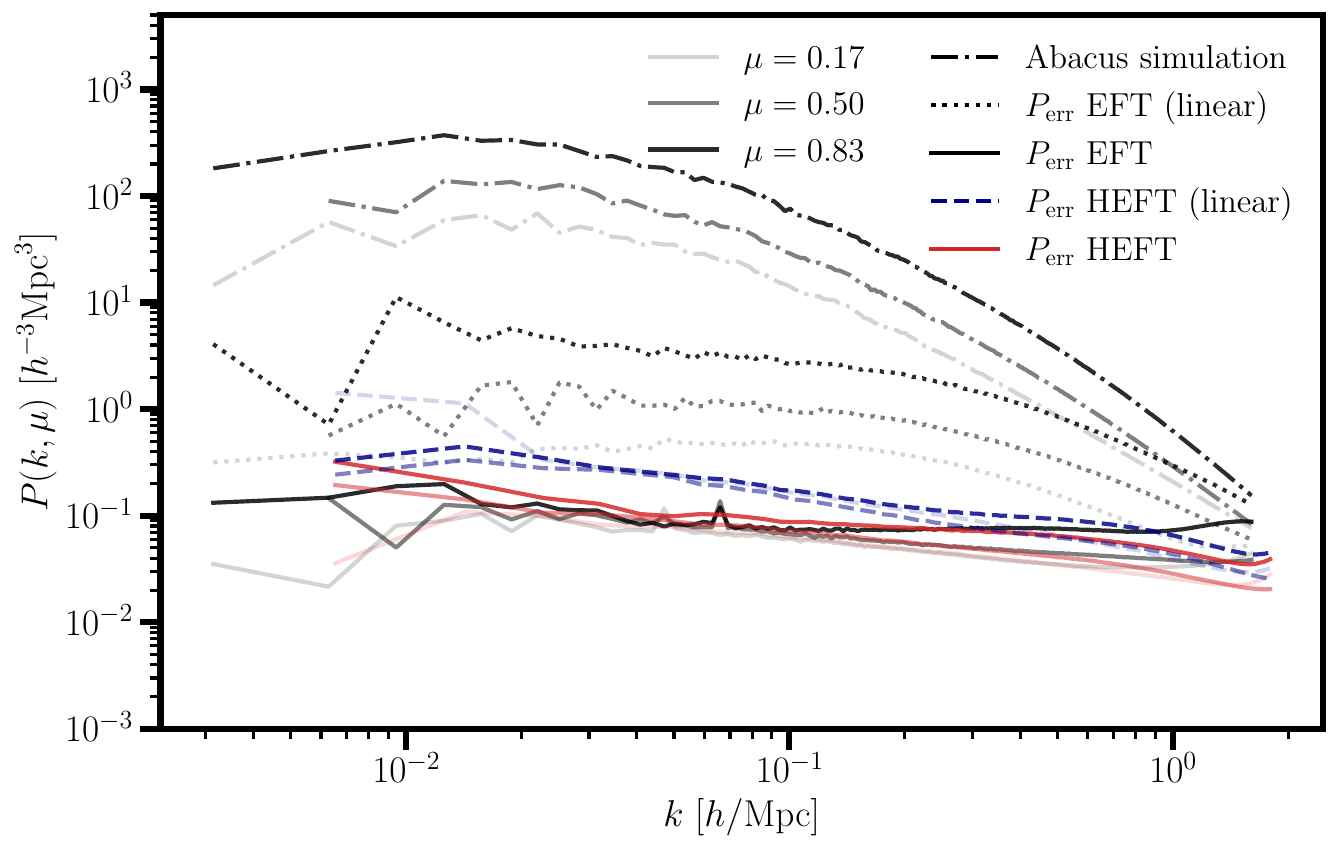}\hfill
\caption{\textbf{Error Power Spectrum:} Comparison of the measured
  power spectrum on the \abacus simulation (dash-dotted black line) for
  model one (left panel) and model three (right panel) to the error
  power spectra $P_{\mathrm{err}}(k,\mu) \equiv \langle
  |\delta^{\mathrm{truth}}_F - \delta^{\mathrm{model}}_F|^2 \rangle$
  obtained from the EFT  results (linear model as dotted black
  and cubic model as solid black line;  \cite{deBelsunce:2025bqc})
  to the HEFT results (linear dashed blue, cubic solid red). In each panel,
  the power spectrum is shown in bins of
  Fourier wavenumber $k$ and angle to the line of sight, parametrized
  by $\mu = k_\parallel / k$, with decreasing line intensity for
  decreasing values of $\mu = 0.83,\, 0.50,\, 0.17$. The corresponding
  ratio plots of the power spectrum of the forward modeled field are
  given in Fig.~\ref{fig:Deltapk} and the corresponding
  cross-correlation coefficients in Fig.~\ref{fig:rcc}. Whilst linear theory
  shows a breakdown at all scales emphasizing the need for higher-order
  bias expansions to capture the non-linearities in the simulation,
  these results illustrate the scales at which HEFT
  yields the largest improvements ($k \simgt 0.1 \hMpcinv$) by removing
  the scale and orientation dependence of the error power spectrum. The recovered
error power spectra for the cubic HEFT model are approximately flat, following theoretical EFT predictions \citep{Ivanov:2023yla}.}
\label{fig:pk}
\end{figure*}

In this section we  forward model the \LyaF transmitted flux fraction, $\td_F$, by evolving a set of initial conditions using a hybrid EFT (HEFT) model, introduced in
Sec.~\ref{sec:LyaHEFT}, and fit a set of bias transfer functions to the  \abacus simulations, introduced in
Sec.~\ref{sec:simulations}. We assess the
performance of our forward model in
Sec.~\ref{sec:results_fits} and quantify the agreement between the phases and
amplitudes by comparing the error power spectrum and
cross-correlation coefficient. We investigate in more detail the obtained transfer functions and 3D \Lya forest stochasticity in Sec.~\ref{sec:pkerr_fits}. We use the EFT model developed in Ref.~\citep{deBelsunce:2025bqc} as our reference and compare the obtained HEFT results using a linear and a cubic perturbative model.  Throughout this section we use the
\textsc{AbacusSummit}
$N$-body simulation suite with a fiducial \textit{Planck} 2018
cosmology: $\Omega_b h^2 = 0.02237$,
$\Omega_c h^2 = 0.12$, $h = 0.6736$, $10^9 A_s = 2.0830$, $n_s =
0.9649$, $w_0 = -1$, $w_a = 0$) and will compare two \LyaF models
with different values for the bias parameters (most noticeably in $b_\eta$ representing the amount of RSD in the mocks), stemming from differences in painting
the \LyaF on top of the dark matter fields using a FGPA prescription. These models will follow
the nomenclature of
Ref.~\cite{Hadzhiyska:2023, Abacus_BAO_Lya:2025} and will be referred
to as models `one' and `three' throughout this work as tabulated in
Tab.~\ref{tab:abacus_models}. We remind the reader that model three
is somewhat closer to observational data but with a larger value for
redshift space distortions (RSD) captured by $b_\eta$ (see Tab.~\ref{tab:abacus_models}), model one
represents a Universe with a lower degree of RSD than observational data suggests \cite{DESI_lya_2024, Chabanier:2024knr}.

\subsection{Results of Field-Level Fits}\label{sec:results_fits}

\begin{figure*}
\centering
\includegraphics[width=0.49\linewidth]{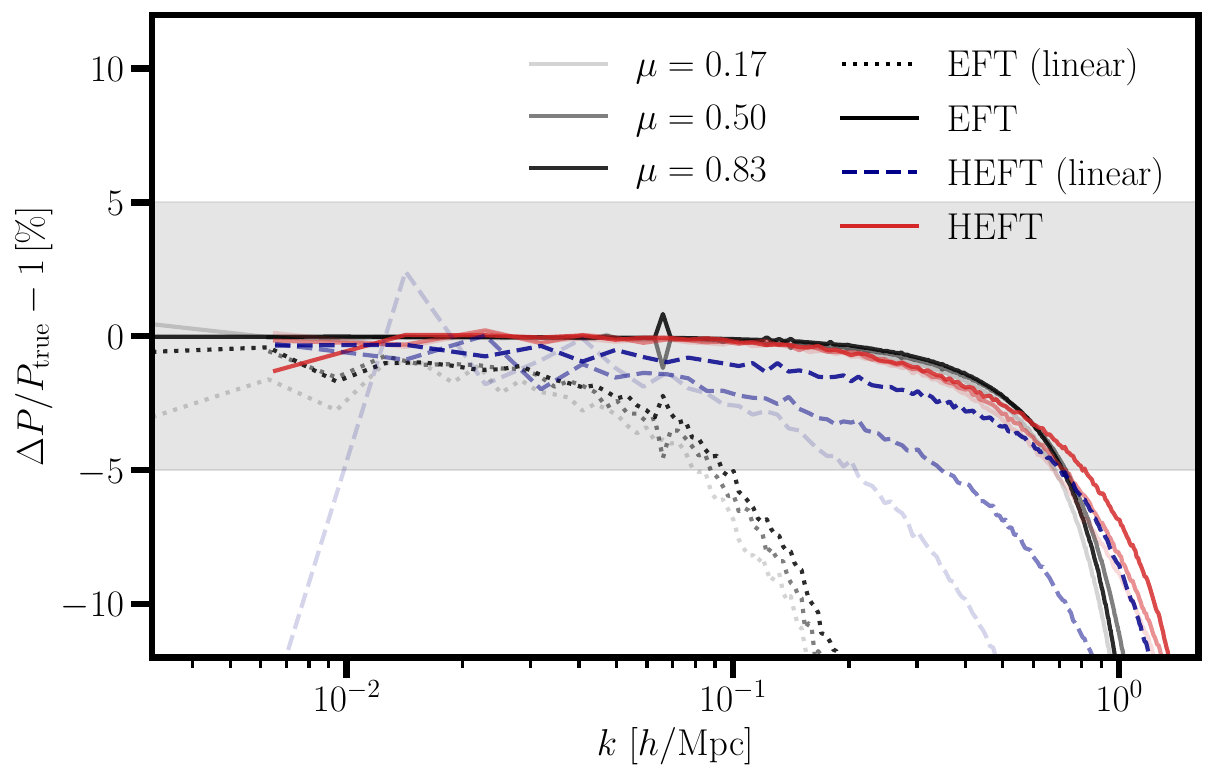}\hfill
\includegraphics[width=0.49\linewidth]{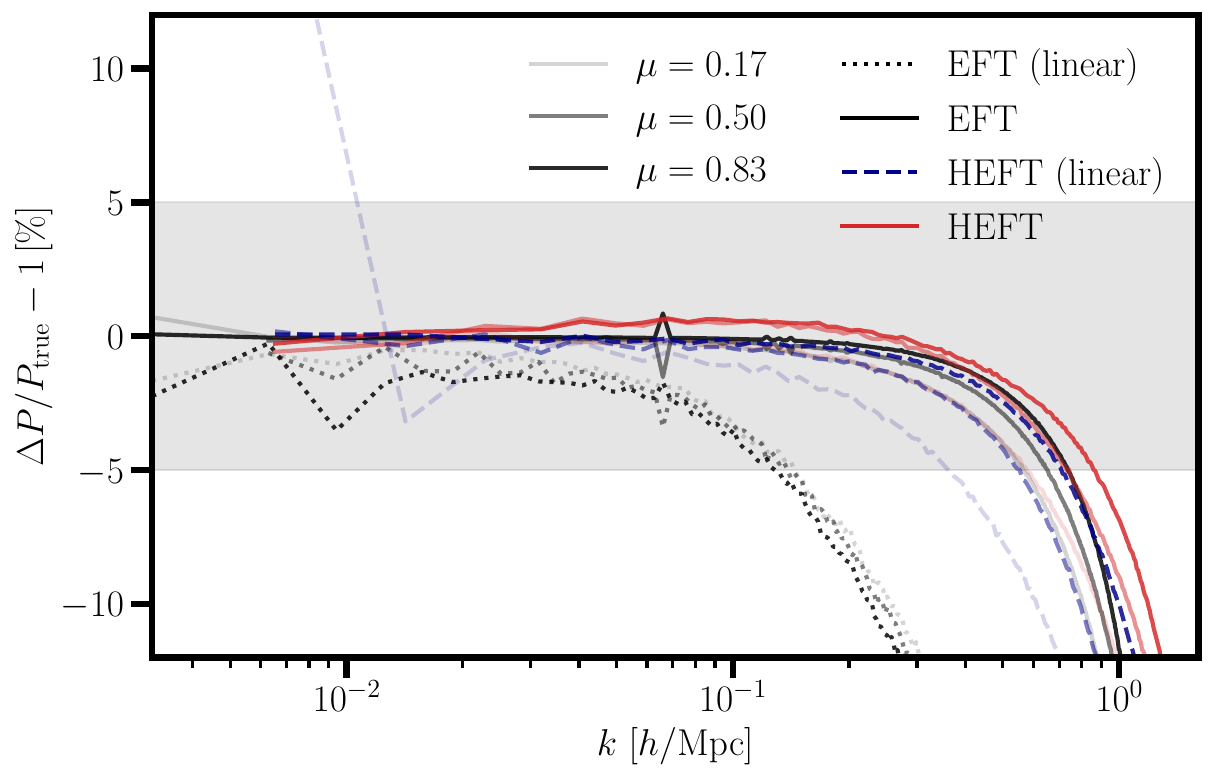}\hfill
\caption{\textbf{Power Spectrum Residuals:} Same as Fig.~\ref{fig:pk}
  for the corresponding power spectrum residuals between the measured
  power spectrum from the \abacus simulation ($P_{\mathrm{true}}$) and
  the power spectra of the forward modeled fields using HEFT with a
  linear (blue) and cubic (red) model compared to the forward model
  using an EFT model (linear as dotted black line, cubic as solid
  black line). We compare two models: one in the
  left panel and three in the right panel and include a 5\% gray error
  band to guide the eye. The power spectrum of the forward model agrees at the 5\% level
  up to $\kmax\simeq 0.8\hMpcinv$ for HEFT, increasing the reach of the
forward model compared to EFT by $\Delta k\approx 0.15 \hMpcinv$.}
\label{fig:Deltapk}
\end{figure*}

\begin{figure*}
\centering
\includegraphics[width=0.49\linewidth]{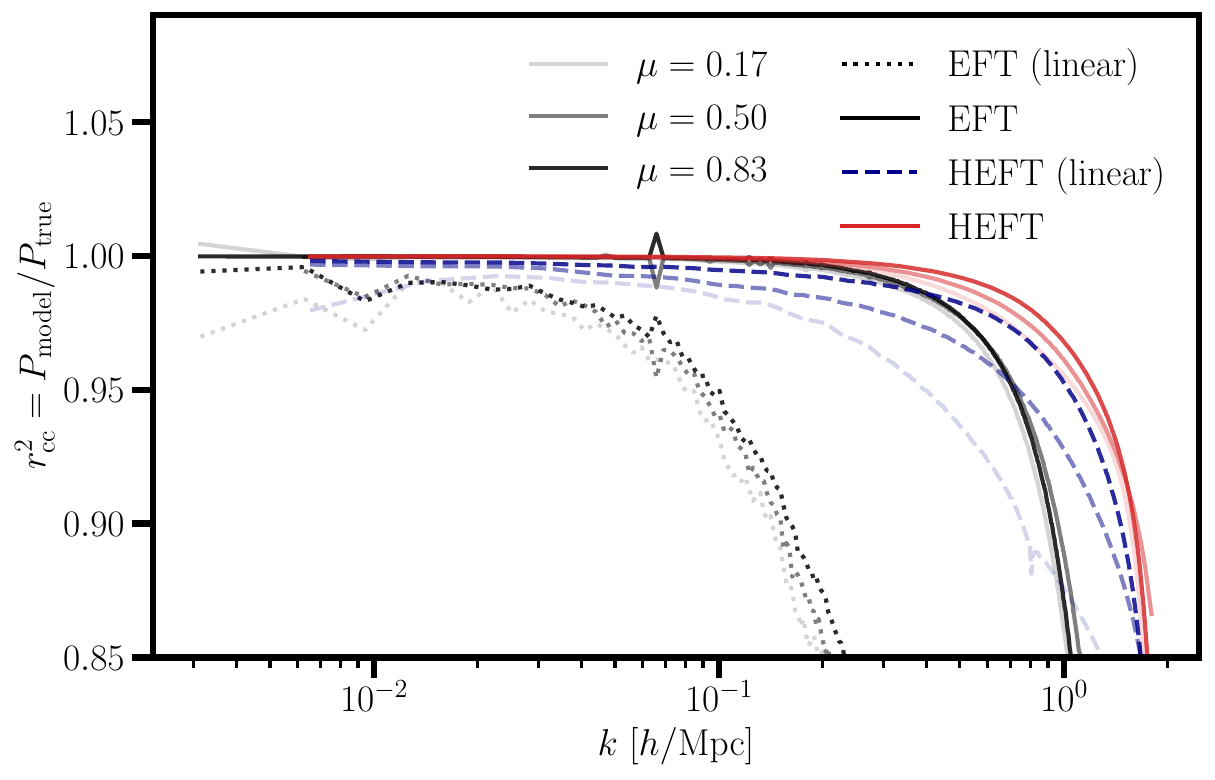}\hfill
\includegraphics[width=0.49\linewidth]{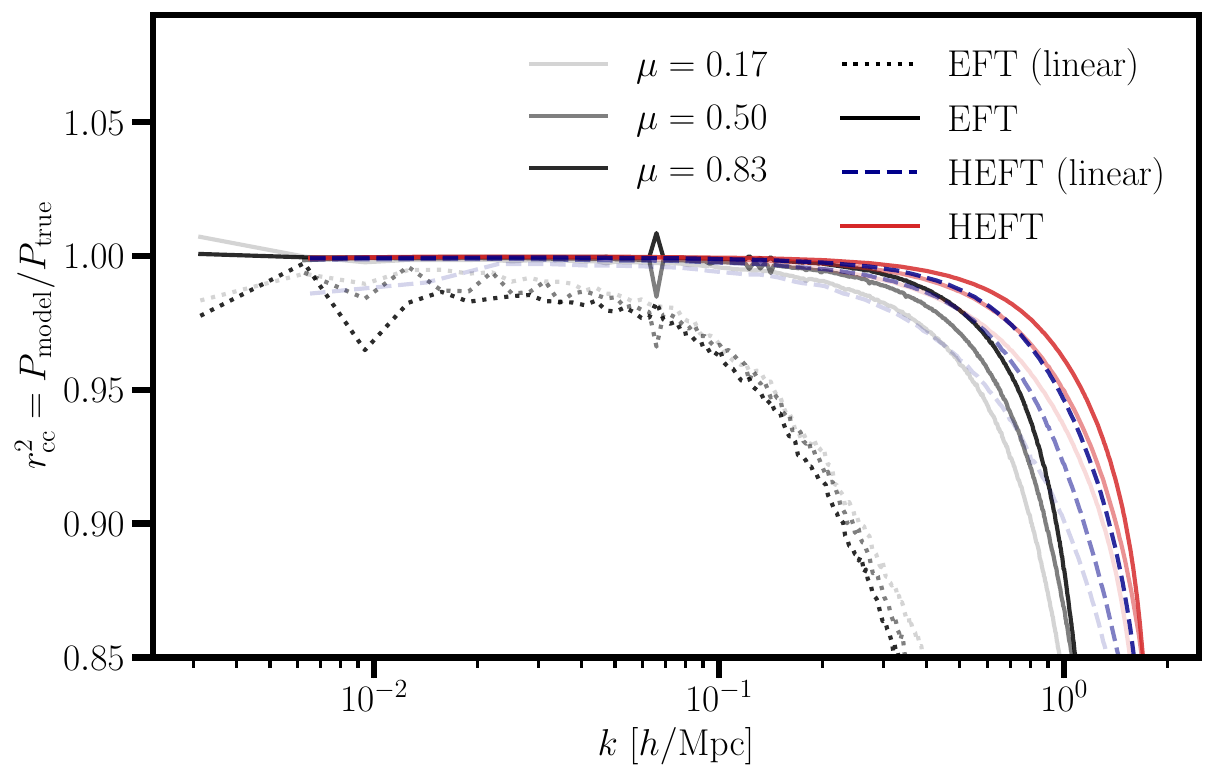}\hfill
\caption{\textbf{Cross-Correlation Coefficient:} Same as
  Fig.~\ref{fig:Deltapk} for the corresponding cross-correlation
  factor $r_{cc}(\delta_F^{\rm truth},\delta^{\rm model}_F)=\langle
  \delta^{\rm model}_F(\k)
  [\delta_F^{\rm truth}(\k)]^*
  \rangle/
  \langle
  (|\delta_F^{\rm truth}(\k)|^2\rangle
    \langle
  |\delta_F^{\rm model}(\k)|^2\rangle)^{1/2}$ between the simulation
  and the forward modeled field for model one (left panel) and model
  three (right panel) illustrating a small and large 
  $b_\eta$ bias parameter, respectively. Following baseline expectation, HEFT improves
  the reach of the cross-correlation coefficient for both models down to
scales of $k\sim 1 \hMpcinv$.}
\label{fig:rcc}
\end{figure*}

In Fig.~\ref{fig:pk}, we compare the measured power spectrum from the \abacus simulation (dash-dotted black line) with those obtained from forward-modeled fields using both the EFT (linear dotted and cubic solid black lines) and HEFT frameworks (linear model in blue, cubic model in red). The corresponding ratios between the simulation and model power spectra are shown in Fig.~\ref{fig:Deltapk}. Results are presented in three angular bins centered at $\mu={0.17,0.5,0.83}$, and for Fourier wavenumber $k$, with the left (right) panel corresponding to a Universe with a low (high) redshift-space distortion parameter, see Tab.~\ref{tab:abacus_models}.

To quantify model performance, we focus on the error power spectrum,
$P_{\mathrm{err}}(k,\mu) \equiv \langle |\delta^{\mathrm{truth}}_F - \delta^{\mathrm{model}}_F|^2 \rangle $
which captures discrepancies in both phase and amplitude between the modeled and true fields. As shown in the left panel of Fig.~\ref{fig:pk}, both linear theory models deviate significantly from the simulation even on large scales, leading to a strongly scale- and orientation-dependent error spectrum. In contrast, linear-theory HEFT already systematically reduces $P_{\mathrm{err}}$ on smaller scales, consistent with expectations that using the non-linear instead of Zel'dovich displacements captures (some of the) non-linear physics. However, it is interesting to note that EFT linear theory for model three exhibits a scale- and orientation dependent error power spectrum already at the largest scales, which HEFT improves over a wide range of scales. This highlights the regime where HEFT most effectively extends the validity of the field-level model.

Using the cubic model (red solid line for HEFT), we find a comparable amplitude to the cubic EFT model (black solid line) for the error power spectra. On large scales, both approaches yield consistent amplitudes, while HEFT further suppresses the scale dependence of the error power spectrum beyond $k \sim 0.3\hMpcinv$. Incorporating particle displacements significantly improves the forward model performance on small scales, producing an approximately constant error power spectrum down to $k \simlt 1\hMpcinv$. The up-turn of the error power spectrum at $k\approx 1.8\hMpcinv$ stems from the chosen grid resolution. 

A similar trend is observed for the high-$b_\eta$ case shown in the right panel. HEFT notably improves the $\mu$-dependence of the linear theory error spectrum. The cubic HEFT model (red line) reduces the overall amplitude and removes most of the orientation dependence of $P_{\mathrm{err}}$, but retains a residual scale dependence -- an overall tilt -- indicating that the model struggles to accurately reconstruct the field in the presence of strong redshift-space distortions.

This suggests that, for cosmological analyses employing HEFT (e.g., via an emulator), the model’s validity range can be extended by marginalizing over a constant “shot-noise”-like term to account for this residual offset. Doing so allows cosmological information to be extracted robustly up to $\kmax \simlt 1\hMpcinv$, with the remaining modeling uncertainty dominated by the small residual scale dependence visible in Fig.~\ref{fig:pk}.

In Fig.~\ref{fig:Deltapk}, we present the ratio between the power spectra of the forward-modeled fields and those measured from the simulations for both EFT and HEFT. On large scales, HEFT introduces percent-level noise due to the limited number of available modes,\footnote{Ref.~\cite{2021JCAP...09..020H} constructed a hybrid HEFT and LPT-based emulator for cosmological data analysis to address this.} but this effect is purely statistical. On smaller scales, HEFT substantially extends the range of validity of the forward model, achieving 5\% (1\%) agreement with the simulation power spectrum up to $k \simlt 1\hMpcinv$ ($k \simlt 0.3\hMpcinv$).

In Fig.~\ref{fig:rcc}, we show the cross-correlation coefficient,
$r_{\mathrm{cc}}(k,\mu)$, which is related to the error power spectrum by  $P_{\mathrm{err}} = P_{\mathrm{truth}}(1 - r_{\mathrm{cc}}^2)$
The linear theory model performs adequately on large scales, with deviations at the $\sim3\%$ level, but its accuracy rapidly degrades beyond $k \sim 0.1\,h\,\mathrm{Mpc}^{-1}$, exhibiting a pronounced $\mu$-dependence for both cases. In contrast, the cubic HEFT model achieves $r_{\mathrm{cc}} > 0.95$ up to $k \sim 1.1\hMpcinv$ for model one (and $k \sim 1.0\hMpcinv$ for model three), demonstrating that incorporating nonlinear particle displacements within the HEFT framework yields a substantial improvement in field-level reconstruction accuracy.


The extension
of the validity range of the EFT model 
is expected to have 
a significant impact
on the parameter constraints from the 
3D forest correlations.
One can use the ratio of 
numbers of 
quasi-linear modes in the 
analysis (i.e. the cumulative power spectrum signal-to-noise) as an estimate 
for parameter variance 
reduction. 
Increasing $k_{\rm max}$
from $0.8~h\text{Mpc}^{-1}$ to 
$1~h\text{Mpc}^{-1}$ increases the 
signal-to-noise ratio by 
a factor of $\approx 2$,
equivalent to a 
twofold increase 
in the observed volume,
or a reduction of up to $\approx 40\%$ 
of cosmological parameter errorbars.

\subsection{Transfer Function Fits} \label{sec:pkerr_fits}
Ref.~\cite{deBelsunce:2025bqc} presented the first estimates of the three-dimensional \LyaF\ stochasticity, which encapsulates the background of nonlinear structure formation. This stochastic component represents an irreducible error floor uncorrelated with the cosmological initial conditions. While EFT provides a first-principles description of the deterministic part of the field, its stochastic contribution is generally treated phenomenologically through a simple power-law momentum expansion. Notably, the HEFT error power spectrum reproduces the EFT theoretical expectation -- displaying only a weak scale dependence~\cite{Ivanov:2023yla} -- analogous to the one-halo term~\cite{Irsic:2018hhg}.

In this section, we examine the error power spectrum and corresponding transfer functions in greater detail. We first fit $P_{\mathrm{err}}(k,\mu)$ using the functional form
\begin{equation} \label{eq:Perr_tf}
P_{\text{err}}(k, \mu) = a_0 + a_2 k^2 + a_3 k^3 + a_4 k^4 + \sum_{i=2,4} a_{ii} (k \mu)^i\,,
\end{equation}
with best-fit parameters for both models listed in Tab.~\ref{tab:abacus_perr_results}. The fit is obtained via a weighted least-squares procedure up to $\kmax = 1\hMpcinv$, jointly across all $\mu$ bins and with each $k$-bin weighted by $k$ to account for mode discreteness.\footnote{We have verified that varying the maximum wavenumber by $\Delta k = 0.2\hMpcinv$ does not affect our conclusions.} Whilst EFT does not allow odd powers of $\mu$, we include terms linear in $k$ to account for higher-order loop corrections. 

\begin{figure}
\centering
\includegraphics[width=\linewidth]{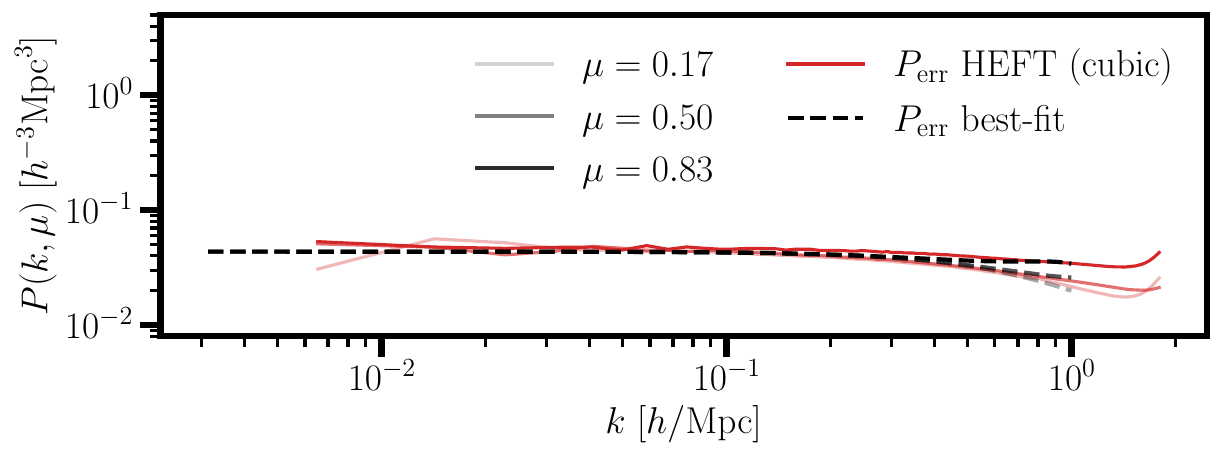}\\
\includegraphics[width=\linewidth]{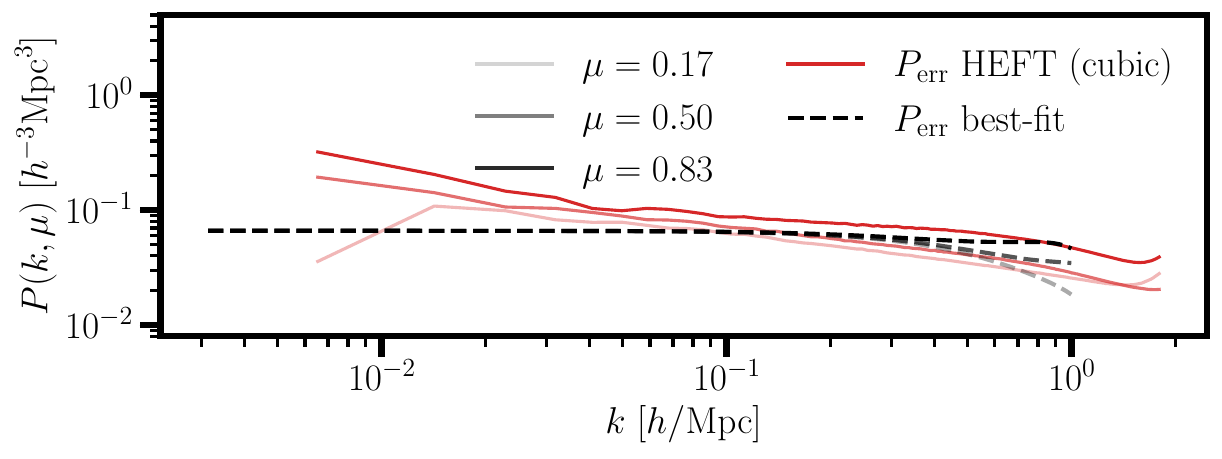}
\caption{\textbf{Noise Power Fits:} We fit the polynomial given in Eq.~\eqref{eq:Perr_tf} to the error power spectrum for model one (three) with low (large) $b_\eta$ bias parameter value in the top (bottom) panel. In a low-$b_\eta$ Universe, the scale-dependence is completely removed (model one), whereas model three with a large $b_\eta$ parameter exhibits a residual scale-dependence on large scales and a strongly suppressed one on intermediate to small scales. The comparison of the fits to the measured error power spectra emphasizes the performance gains obtained by HEFT on scales beyond $k\simgt 0.2 \hMpcinv$. }
\label{fig:pkerr_fits}
\end{figure}

As shown in Tab.~\ref{tab:abacus_perr_results}, the overall amplitude is lower by one third for model one (the low-$b_\eta$ scenario). The fits, compared to the measured error power spectra in Fig.~\ref{fig:pkerr_fits}, demonstrate that HEFT almost entirely removes both the scale and orientation dependence for model one, substantially improving the small-scale behavior. Although HEFT continues to outperform EFT for model three, a residual scale dependence -- approximately an order of magnitude larger -- remains for the large-$b_\eta$ scenario. Nevertheless, at small to intermediate scales this dependence is largely mitigated, suggesting that a hybrid emulator approach is well-motivated: employing EFT on large scales (where HEFT may encounter numerical limitations) and switching to HEFT for $k \simgt 0.2 \hMpcinv$.

\begin{table}
\centering
\begin{tabular}{ccccccc}
\hline\hline
M. &$a_0$ & $a_2$ & $a_3$ & $a_4$ & $a_{22}$ &  $a_{44}$ \\
\hline
I   &$\phantom{-}0.041$ & $-0.047$ & $\phantom{-}0.365$ & $-0.009$ & $\phantom{-}0.018$ &  $-0.003$ \\
III & $\phantom{-}0.060$ & $-0.087$ & $\phantom{-}0.071$ & $-0.018$ & $\phantom{-}0.029$ &  $-0.010$ \\
\hline\hline
\end{tabular}
\caption{\textbf{Noise Power Best-Fit Values:} Coefficients for the polynomial fit to $P_{\rm err}$ given in Eq.~\eqref{eq:Perr_tf} ($a_n, a_{nn}$ are in units of ~$[\Mpch]^{3+n}$) for the \abacus simulation model one (top row) and model three (bottom row), shown in Fig.~\ref{fig:pk}. The fits are obtained via a weighted least-squares procedure up to $\kmax = 1\hMpcinv$ using all $\mu$ bins jointly and weighting each $k$-bin by its number of modes.}
 \label{tab:abacus_perr_results}
\end{table}

To characterize the behavior of the transfer functions, we adopt the following functional form for their scale and angular dependence. We model each transfer function $\beta(k,\mu)$ (where $\beta$ denotes the polynomial fits to the bias transfer functions) using the Ansatz
\begin{align} \label{eq:beta_tf}
\beta_i(k,\mu) &= c_0 + c_{01}\mu^2 \\
&\quad + \left(c_1 + c_{12}\mu^2 + c_{14}\mu^4\right) k \nonumber \\
&\quad + \left(c_4 k^2 + c_{22}\mu^2 + c_{44}k^2\mu^4\right) k^2\,, \nonumber
\end{align}
where $\beta_{cb}$ is the transfer function for the non-linear RSD field and the remaining nine transfer functions are for each bias parameter in the cubic bias expansion from Sec.~\ref{sec:HEFT_operators}. Note that this definition differs from the parametrization used in Refs.~\cite{Schmittfull:2020trd, Obuljen:2022cjo} and provides additional flexibility to capture the angular dependence of the transfer functions, particularly for $k \to 0$. We apply the same weighted least-squares fitting procedure as for the error power spectrum. 

\begin{figure*}
    \centering
    \includegraphics[width=1\linewidth]{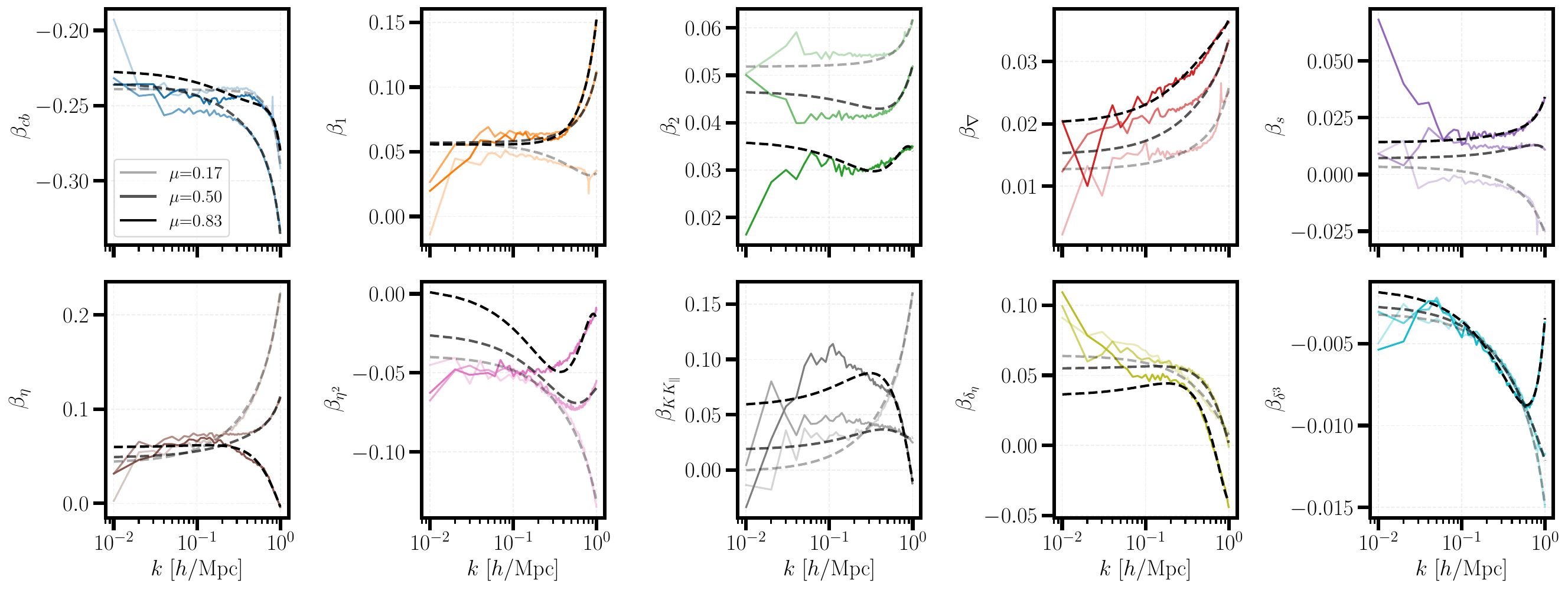}\\
    \includegraphics[width=1\linewidth]{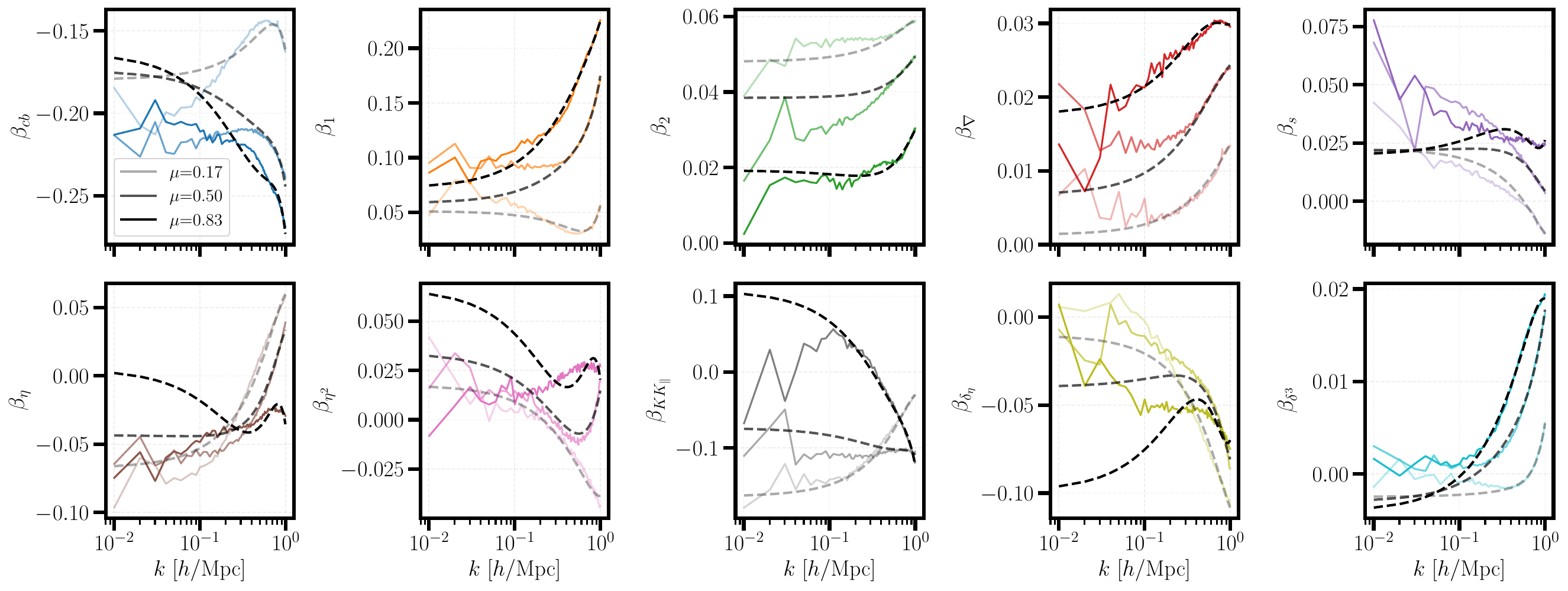}
    \caption{\textbf{Transfer Function Measurements:} Best-fit transfer functions $\beta_i(k,\mu)$ for the cubic HEFT model obtained from fits to the \abacus simulation for model one (top panel) and model three (bottom panel). The corresponding polynomial model for the transfer functions $\beta(k,\mu)$ is given in Eq.~\eqref{eq:beta_tf} which reduces to the Kaiser model in the low-$k$ limit. The large fluctuations in the first two $k$-bins stem from the small number of available modes. Weighting each bin by its number of $k$ modes down weights very noisy bins. The coefficients are tabulated in Tab.~\ref{tab:beta_parameters}. }
    \label{fig:transfer_func}
\end{figure*}

The measured and best-fit transfer functions are shown in Fig.~\ref{fig:transfer_func} with the best-fit polynomial parameters tabulated in Tab.~\ref{tab:beta_parameters} for model one (three) in the top (bottom) panel. Qualitatively both sets of transfer function agree with each other and with increasing RSD, we find a slightly increased $\mu$ dependence of the transfer functions. This figure illustrates that the transfer functions can be approximated by smooth functional forms. A powerful test of this framework, would be to theoretically predict the shapes of the transfer functions (see, e.g.,~\cite{Schmittfull:2018yuk,Schmittfull:2020trd,Ivanov:2024hgq,Ivanov:2024jtl}) -- we present the results in a companion paper \cite{Belsunce_field_long:25}. 

\begin{table*}
\centering
  \setlength{\tabcolsep}{8pt}
\begin{tabular}{l|cccccccc}
\hline\hline
TF & $c_0$ & $c_{01}$ & $c_1$ & $c_{12}$ & $c_{14}$ & $c_4$ & $c_{22}$ & $c_{44}$ \\
\hline
\textbf{Model I} &&&&&&&&\\
$\beta_{cb}$ & $-0.239$ & $\phantom{-}0.018$ & $\phantom{-}0.009$ & $-0.535$ & $\phantom{-}0.561$ & $-0.050$ & $\phantom{-}0.164$ & $-0.054$  \\
$\beta_{1}$ & \phantom{-}0.057 & -0.002 & -0.051 & \phantom{-}0.348 & -0.411 & \phantom{-}0.015 & \phantom{-}0.115 & \phantom{-}0.022 \\
$\beta_{2}$ & \phantom{-}0.052 & -0.023 & \phantom{-}0.006 & -0.116 & \phantom{-}0.072 & \phantom{-}0.005 & \phantom{-}0.094 & -0.068 \\
$\beta_{\nabla}$ & \phantom{-}0.012 & \phantom{-}0.011 & \phantom{-}0.008 & \phantom{-}0.070 & -0.049 & \phantom{-}0.004 & -0.033 & \phantom{-}0.006 \\
$\beta_{s}$ & \phantom{-}0.003 & \phantom{-}0.016 & -0.028 & \phantom{-}0.215 & -0.221 & -0.008 & -0.011 & \phantom{-}0.043 \\
$\beta_{\eta}$ & \phantom{-}0.042 & \phantom{-}0.025 & \phantom{-}0.161 & -0.441 & \phantom{-}0.370 & \phantom{-}0.038 & -0.183 & -0.016 \\
$\beta_{\eta^2}$ & -0.041 & \phantom{-}0.065 & -0.088 & -0.315 & -0.002 & -0.014 & \phantom{-}0.671 & -0.341 \\
$\beta_{(KK)_{\parallel}}$ & -0.004 & \phantom{-}0.088 & \phantom{-}0.152 & -0.418 & \phantom{-}0.697 & \phantom{-}0.035 & -0.477 & \phantom{-}0.060 \\
$\beta_{\delta\eta}$ & \phantom{-}0.066 & -0.043 & -0.066 & \phantom{-}0.446 & -0.300 & \phantom{-}0.008 & -0.400 & \phantom{-}0.199 \\
$\beta_{\delta^3}$ & -0.003 & \phantom{-}0.002 & -0.010 & -0.012 & -0.006 & -0.002 & \phantom{-}0.019 & \phantom{-}0.017 \\
\hline
\textbf{Model III} &&&&&&&&\\
$\beta_{cb}$ & -0.180 & \phantom{-}0.024 & \phantom{-}0.083 & -0.966 & \phantom{-}0.635 & -0.051 & \phantom{-}0.438 & -0.171 \\
$\beta_{1}$ & \phantom{-}0.050 & \phantom{-}0.031 & -0.061 & \phantom{-}0.828 & -0.578 & \phantom{-}0.048 & -0.147 & -0.057 \\
$\beta_{2}$ & \phantom{-}0.049 & -0.043 & \phantom{-}0.014 & -0.047 & \phantom{-}0.011 & -0.003 & \phantom{-}0.047 & -0.011 \\
$\beta_{\nabla}$ & \phantom{-}0.001 & \phantom{-}0.024 & \phantom{-}0.012 & \phantom{-}0.096 & -0.075 & -0.001 & -0.057 & \phantom{-}0.021 \\
$\beta_{s}$ & \phantom{-}0.023 & -0.004 & -0.046 & \phantom{-}0.271 & -0.147 & \phantom{-}0.007 & -0.170 & \phantom{-}0.098 \\
$\beta_{\eta}$ & -0.071 & \phantom{-}0.108 & \phantom{-}0.166 & -0.818 & \phantom{-}0.270 & -0.033 & \phantom{-}0.607 & -0.329 \\
$\beta_{\eta^2}$ & \phantom{-}0.016 & \phantom{-}0.073 & -0.079 & -0.285 & \phantom{-}0.020 & \phantom{-}0.015 & \phantom{-}0.551 & -0.368 \\
$\beta_{(KK)_{\parallel}}$ & -0.176 & \phantom{-}0.407 & \phantom{-}0.194 & -1.381 & \phantom{-}0.672 & -0.035 & \phantom{-}0.480 & -0.184 \\
$\beta_{\delta\eta}$ & -0.007 & -0.133 & -0.127 & \phantom{-}0.840 & -0.382 & \phantom{-}0.020 & -0.544 & \phantom{-}0.236 \\
$\beta_{\delta^3}$ & -0.002 & -0.002 & -0.000 & \phantom{-}0.094 & -0.056 & \phantom{-}0.006 & -0.015 & -0.024 \\
\hline\hline
\end{tabular}
\caption{\textbf{Best-Fit Transfer Function Values:} Best-fit parameters for the transfer function (TF) model, $\beta(k,\mu)$, given in Eq.~\eqref{eq:beta_tf} and illustrated in Fig.~\ref{fig:transfer_func} obtained from the \abacus simulation for model one (top half) and model three (bottom half). Each wavenumber bin is weighted by $k$ to down weight small scale modes with a cut off at $k_{\rm max}=1\hMpcinv$. The coefficients $c_n, c_{nm}$ are given in units of ~$[\Mpch]^{n}$.}
\label{tab:beta_parameters}
\end{table*}

\begin{figure*}
    \centering
    \includegraphics[width=1\linewidth]{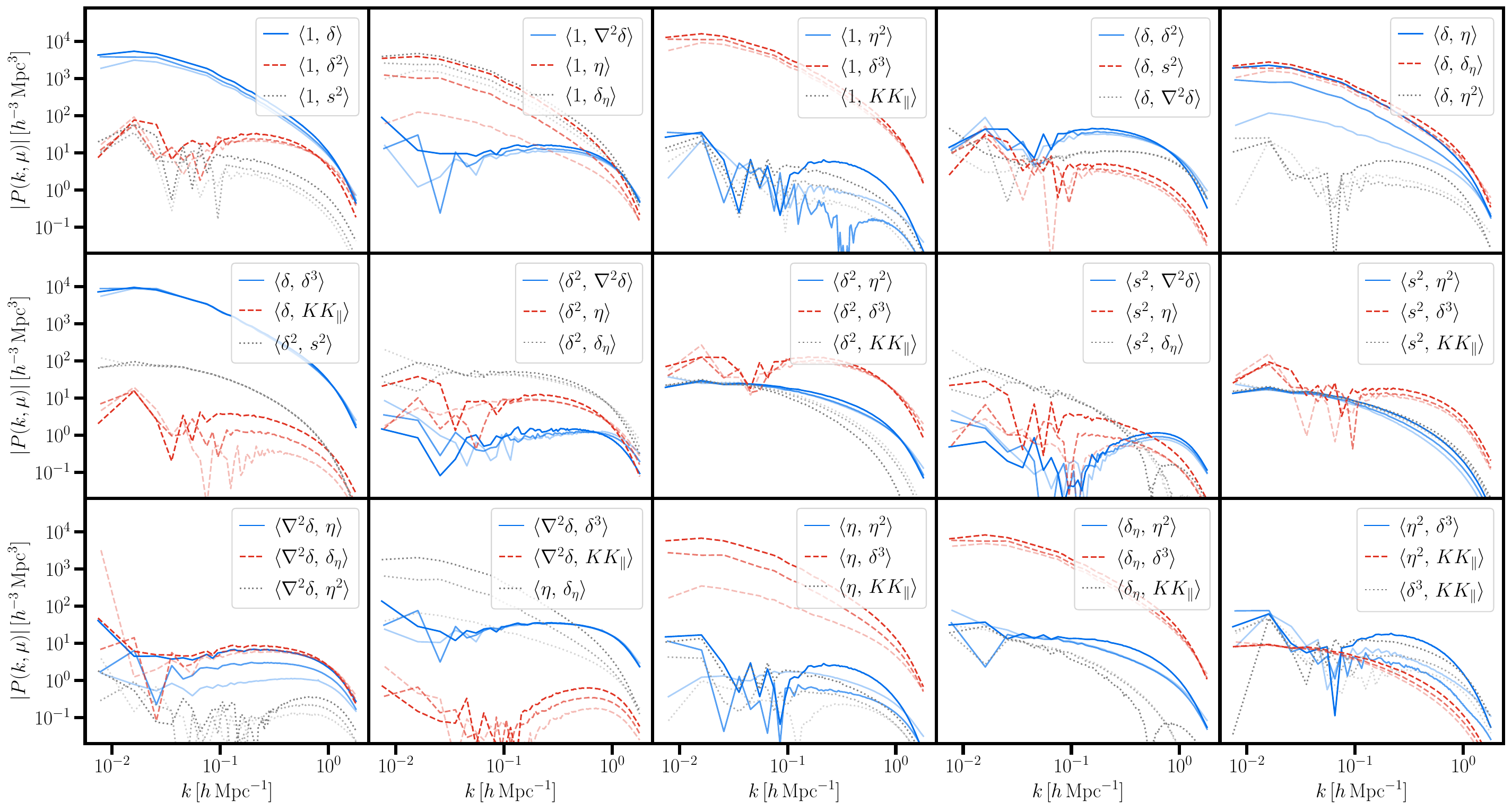}
    \caption{\textbf{Cross-Spectra of HEFT Fields:} Basis spectra at $z=2.5$ as a function of Fourier wavenumber $k$ and three angular bins centered at $\mu=0.17,\, 0.50,\, 0.83$. We plot the absolute value of each field and find that the templates stemming from $1$, $\td$, $\td^3$ as well as the velocity fields $\eta$ and $\td_{\eta}$ show some degeneracy and dominate on large scales.
    }
    \label{fig:Cross_pk_fields}
\end{figure*}

In Fig.~\ref{fig:Cross_pk_fields} we show the basis cross spectra of the different advected fields extracted from the simulations given in Eq.~\eqref{eq:bias_expansion}. We plot all of the 45 combinations of the fields for each $(k,\mu)$-bin used in the fits. Whilst the templates stemming from $1$, $\td$, $\td^3$ and from the velocity fields $\eta$ and $\td_{\eta}$ have similar shapes, they dominate the final power spectrum. The noise in the cross spectra (in particular, combinations with the tidal tensor) are visible in Fig.~\ref{fig:pk}. To reduce the noise in the final field, we advocate using a similar approach chosen in Ref.~\cite{2020MNRAS.492.5754M} to use an analytic model (here: EFT field-level fits) on large scales with a HEFT model on smaller scales. 

\section{Summary and Conclusion} \label{sec:conclusion}
The \Lya forest -- as measured by the currently observing Dark Energy
Spectroscopic Instrument (DESI) -- is a pristine tracer of
large-scale structure at Mpc scales and below at high redshift
($2\leq z \leq 4$). The \Lya forest has four main advantages compared
to traditional galaxy surveys: First, it probes a significantly
larger cosmological volume on the past lightcone which remains
inaccessible to galaxy surveys until Stage-V spectroscopy
\cite{2022arXiv220903585S}. Second, given its high redshift, the
\LyaF probes more quasi-linear modes probing fundamental physics,
extending the reach of perturbative techniques \cite{Chen:2021,Ivanov:2023yla,
deBelsunce:2025edy}. Third, hydrodynamical simulations of the \Lya forest
down to $\sim$kpc scales enable precision calibration of perturbative
models \cite{McDonald:2001, Arinyo-i-Prats:2015, deBelsunce:2024rvv,
Chudaykin:2025gsh, Abacus_BAO_Lya:2025}. Fourth, shot-noise
corrections to the \Lya forest vanish towards higher redshifts (in
agreement with the scaling Universe estimate \cite{Ivanov:2023yla,
deBelsunce:2024rvv}), offering additional insights compared to
galaxies or quasars \cite{Chudaykin:2025gsh}. However, simulations encompassing
cosmological volumes and currently employed heuristic
fitting functions need to meet the challenging requirements imposed
by precise \LyaF observations from DESI.

In this work, we expand on the work in Ref.~\cite{deBelsunce:2025bqc} and
combine a perturbative EFT-based forward model with displacements computed with
the $N$-body simulation \abacus, denoted by hybrid effective field
theory (HEFT; \cite{Kokron:2021xgh, Zennaro:2021bwy}). In a first
step, we show in Sec.~\ref{sec:results_fits}, that our HEFT model can
reproduce the input \LyaF simulations which are painted on top
of $N$-body simulations of the \abacus suite up to $k\simlt 1
\hMpcinv$ ($k\simlt 0.3 \hMpcinv$) at the 5\% (1\%) level, at fixed cosmology. This removes the scale-dependence of the
EFT-based forward model and results in a further $k$-reach of the
hybrid model.\footnote{Note that current DESI \Lya full-shape analyses use the correlation function up to a minimum scale of $r=25\hinvMpc$ \cite{Cuceu:2025nvl} corresponding to $k \approx  \pi/25 \approx 0.13 \hMpcinv$ which is the range of validity where the present forward model is accurate at the sub-percent level.}
Based on the cumulative power spectrum signal-to-noise ratio this increase in $\Delta k_{\rm max}\approx 0.2\hMpcinv$ compared to the purely EFT-based approach ($k_{\rm max}^{\rm EFT}\approx 0.8\hMpcinv$ to $k_{\rm max}^{\rm HEFT}\approx 1.0\hMpcinv $) is equivalent to an increase of the observed volume by a factor of two which, in turn, corresponds to a potential reduction of cosmological parameter errorbars by up to $\sim 40\%$ -- emphasizing the need for robust theoretical tools in the context of cosmological data analysis of current Stage-IV surveys. 
Further, the HEFT technique can be used to generate a \LyaF
field in less than one hour on the Perlmutter computer at NERSC (using one AMD Milan CPU) with a volume of $2\hinvGpc$ with a
DESI-simulation resolution of  $\sim 2.5\hinvMpc$ -- suitable to validate cosmological
inference pipelines (see, e.g.,~\cite{Ramirez-Perez:2021cpq, Abacus_BAO_Lya:2025}).

To enable the fast cosmological analysis of the \LyaF, we advocate
developing an emulator
for summary statistics such as the 3D power spectrum
\cite{deBelsunce:2024knf}, correlation function \cite{Cuceu:2025nvl},
or compressed 3D bispectrum \cite{deBelsunce:2025edy} calibrated on a
suite of \abacus
simulations. Schematically, one would first read in all the stored
\abacus particles for each of the simulations
stored with varying cosmologies and fixed phases. Using the same phase for
the initial conditions will yield cosmic variance cancellations when
taking derivatives of the power spectrum templates with respect to
(cosmological) parameters. Then for each simulation snapshot one
repeats the procedure in Ref.~\cite{Hadzhiyska:2023} to plant a
forest on top of the dark matter fields and then we apply the HEFT
procedure of Sec.~\ref{sec:LyaHEFT} to compute each of the \Lya
specific fields and fit the obtained transfer functions. 
This amounts to weighting each matter particle per
simulation by the value of the corresponding operator in the initial
conditions at the original Lagrangian coordinates $\qvec$. Second,
auto- and cross-power spectra between the transmitted flux fraction
and the fields need to be computed. The measured auto-
and cross-spectra are then given as a sum over a bank of templates
\begin{align}
P_{Fm}(k)&=\sum_{\gamma\in{\beta_{\cal O}}}\beta_\gamma P_{1\gamma}(k)\,, \\
P_{FF}(k)&=\sum_{\gamma\in{\beta_{\cal O}}}\sum_{\delta\in{\beta_{\cal
O}}}\beta_\gamma \beta_\delta P_{\gamma\delta}(k)\,,
\end{align}
where ${\beta_{\cal O}}$ are the quadratic bias operators given in
Eq.~\eqref{eq:bias_expansion}, $P_{\gamma\delta}(k)$ the cross-spectra
of the advected fields $\gamma$ and $\delta$ with corresponding bias
coefficients $\beta_{\gamma,\, \delta}$, and the non-linear power spectrum
is denoted by $P_{11}(k)$ (with a fixed bias parameter of unity).
Thus, the emulator consists of computing all auto- and cross spectrum
templates for all operators. Whilst these templates are calibrated on
the \abacus simulations which use a simple FGPA prescription to paint
the forest on top the dark matter fields, they capture and resolve
the non-linear BAO shift relevant for analysis of
the \LyaF \citep{deBelsunce:2024rvv}. Additionally, one could extend 
this emulator to include cross-correlations with a generic biased tracer of matter, i.e.~quasars, high-redshift galaxies, or dark matter halos \cite{Chudaykin:2025gsh}. We leave these lines of research
to future work.

\section*{Acknowledgments}
It is our pleasure to thank the \LyaF aficionados Jamie Sullivan and Stephen Chen for
fruitful discussions.

This work is supported by the National Science Foundation under Cooperative Agreement PHY-2019786 (The NSF AI Institute for Artificial Intelligence and Fundamental Interactions, http://iaifi.org/). This research used resources of the National Energy Research
Scientific Computing Center (NERSC), a U.S. Department of Energy
Office of Science User Facility operated under Contract No.~DE–AC02–05CH11231.

\section*{Data Availability}
The simulations used in this work are publicly available.
Instructions for access and download are given at
\url{https://abacussummit.readthedocs.io/en/latest/data-access.html}. Instructions for downloading the \textsc{AbacusSummit} based \LyaF mocks can be found in Ref.~\cite{Hadzhiyska:2023}.

\appendix

\bibliographystyle{aux_files/JHEP}
\bibliography{short.bib, references, BH_refs}

\appendix

\end{document}

%% file: aux_files/mycommands.tex
\makeatletter
\def\p@subsection{}
\makeatother

\usepackage{graphicx,amsmath,amssymb,lipsum,bm}

\usepackage[dvipsnames]{xcolor}
\definecolor{xlinkcolor}{rgb}{0.7752941176470588, 0.22078431372549023, 0.2262745098039215}
\definecolor{BrickRed}{rgb}{0.7752941176470588, 0.22078431372549023, 0.2262745098039215}
\definecolor{xlinkcolor}{HTML}{1c1e94}
\usepackage{booktabs} 
\usepackage[normalem]{ulem}
\usepackage[utf8]{inputenc} 
\usepackage{mathtools}
\usepackage{aux_files/aas_macros}
\usepackage{xspace}

\usepackage{ulem}
\usepackage{diagbox}

\usepackage[dvipsnames]{xcolor}
\usepackage{mathrsfs}

\usepackage[colorlinks=true,citecolor=xlinkcolor,linkcolor=xlinkcolor,urlcolor=xlinkcolor, backref=false,pdfborder={0 0 0}]{hyperref}

\newcommand{\beqa}{\begin{eqnarray}}
\newcommand{\eeqa}{\end{eqnarray}}

\newcommand{\be}{\begin{equation}}
\newcommand{\ee}{\end{equation}}
\newcommand{\beq}{\begin{equation}}
\newcommand{\eeq}{\end{equation}}

\renewcommand\k{{\bm k}}

\def\msunoh{\,h^{-1}{\rm M}_\odot}

\newcommand{\bseq}{\begin{subequations}}
\newcommand{\eseq}{\end{subequations}}

\def\ltsima{$\; \buildrel < \over \sim \;$\xspace}
\def\gtsima{$\; \buildrel > \over \sim \;$\xspace}
\def\simlt{\lower.5ex\hbox{\ltsima}}
\def\simgt{\lower.5ex\hbox{\gtsima}}

\newcommand{\hinvMpc}{\,h^{-1}\, {\rm Mpc}\xspace}

\newcommand{\hMpcinv}{\,h\, {\rm Mpc}^{-1}\xspace}
\newcommand{\hinvGpc}{\,h^{-1}\, {\rm Gpc}\xspace}

\newcommand{\Mpch}{\, h^{-1}\mathrm{Mpc}\, }

\newcommand{\kmax}{\, k_{\rm max}\, }

\newcommand{\vk}{\bm{k}}

\newcommand{\Lya}{Ly-$\alpha$\xspace}
\newcommand{\LyaF}{Ly-$\alpha$ forest\xspace}
\newcommand{\td}{\delta}

\newcommand{\qvec}{\bm{q}}

\def\gsim{\raise0.3ex\hbox{$\;>$\kern-0.75em\raise-1.1ex\hbox{$\sim\;$}}}
\def\lsim{\raise0.3ex\hbox{$\;<$\kern-0.75em\raise-1.1ex\hbox{$\sim\;$}}}

\def\beqn#1{\begin{equation}\label{#1}}
\def\eeqn{\end{equation}}

\def\beqa#1{\begin{eqnarray}\label{#1}}
\def\eeqa{\end{eqnarray}}

\newcommand{\abacus}{\textsc{AbacusSummit}\xspace}

\def\kmax{{k_\text{max}}}

\def\Mpch{h^{-1}{\text{Mpc}}}

\def\Z2{$\mathcal{Z_2}$}

\def\msunoh{\,h^{-1}{\rm M}_\odot}

\newcommand {\ignore}[1]{}

\DeclareRobustCommand{\ion}[2]{%
\relax\ifmmode
\ifx\testbx\f@series
{\bm{#1\,\mathsc{#2}}}\else
{\mathrm{#1\,\mathsc{#2}}}\fi
\else\textup{#1\,{\mdseries\textsc{#2}}}%
\fi}

\newcommand{\MIT}{Center for Theoretical Physics -- a Leinweber Institute, Massachusetts Institute of Technology, Cambridge, MA 02139, USA}
\newcommand{\IAIFI}{The NSF AI Institute for Artificial Intelligence and Fundamental Interactions, Cambridge, MA 02139, USA}

\newcommand{\Kavli}{Institute of Astronomy \& Kavli Institute for Cosmology, University of Cambridge, Madingley Road, Cambridge CB3 OHA, UK}